\documentclass[a4paper,12pt,twocolumn]{article}
\usepackage[T1]{fontenc}
\usepackage{comment}
\usepackage{amsfonts}
\usepackage{ae,aecompl}
\usepackage{amsmath}
\usepackage{amssymb}
\usepackage{amsthm}
\usepackage[pdftex]{graphicx}
\usepackage{pstricks}
\usepackage{tikz}
\usepackage{epstopdf}
\usepackage[small]{caption}   
\usepackage{color}
\usepackage{enumerate} 
\usepackage{floatrow}
\usepackage{hyperref}
\usepackage{mathdots}
\usepackage{hyperref}
\hypersetup{
    colorlinks=true,
    linkcolor=blue,
    filecolor=magenta,      
    urlcolor=blue,
}

\definecolor{lightgray}{gray}{0.8}

\usepackage{siunitx}

\title{Dynamic 3D Tomographic X-ray Data of Ladybug}
\author{J. Juurakko,\footnote{Department of Automation and Electrical Engineering, Aalto University, Finland (juliaana.juurakko@aalto.fi)}
\ Z. Purisha,\footnote{Department of Automation and Electrical Engineering, Aalto University, Finland (zenith.purisha@aalto.fi)}  
\ and S. S\"arkk\"a\footnote{Department of Automation and Electrical Engineering, Aalto University, Finland (simo.sarkka@aalto.fi)}}

\date{}
\begin{document}
\maketitle
\abstract{This is the documentation of the 3D dynamic tomographic X-ray (CT) data of a ladybug. 
The open data set is available \href{https://zenodo.org/record/3375488#.XV_T3vxS9oA}{here} 
and can be freely used for scientific purposes with appropriate references to the data and to this document in \url{http://arxiv.org/}. 
}

\section{Introduction}
The main idea of this project was to create dynamic 3D CT data from the object (a ladybug). 3D data is interesting because in medical CT the objects are usually interesting in all three dimensions \cite{Aoki1115}. The dynamic of this dataset was also interesting because when CT measurement is executed for a living object, such as a human body or for a part of it, it is never totally static \cite{Huang2009}. Data set could be used also for limited data dynamic tomography by choosing the corresponding data. The sparse data is useful for reducing both the radiation dose of the patient and the measurement time. It can be also possible that the scanning angle is limited. \cite{sidky2008image}

In this documentation, CT measurements were executed with 360 projections and the eight different measurements of the ladybug were done. The different starting positions are shown in Figure \ref{fig:slices}.

\bigskip
\begin{figure}[h]
\includegraphics[width=7cm]{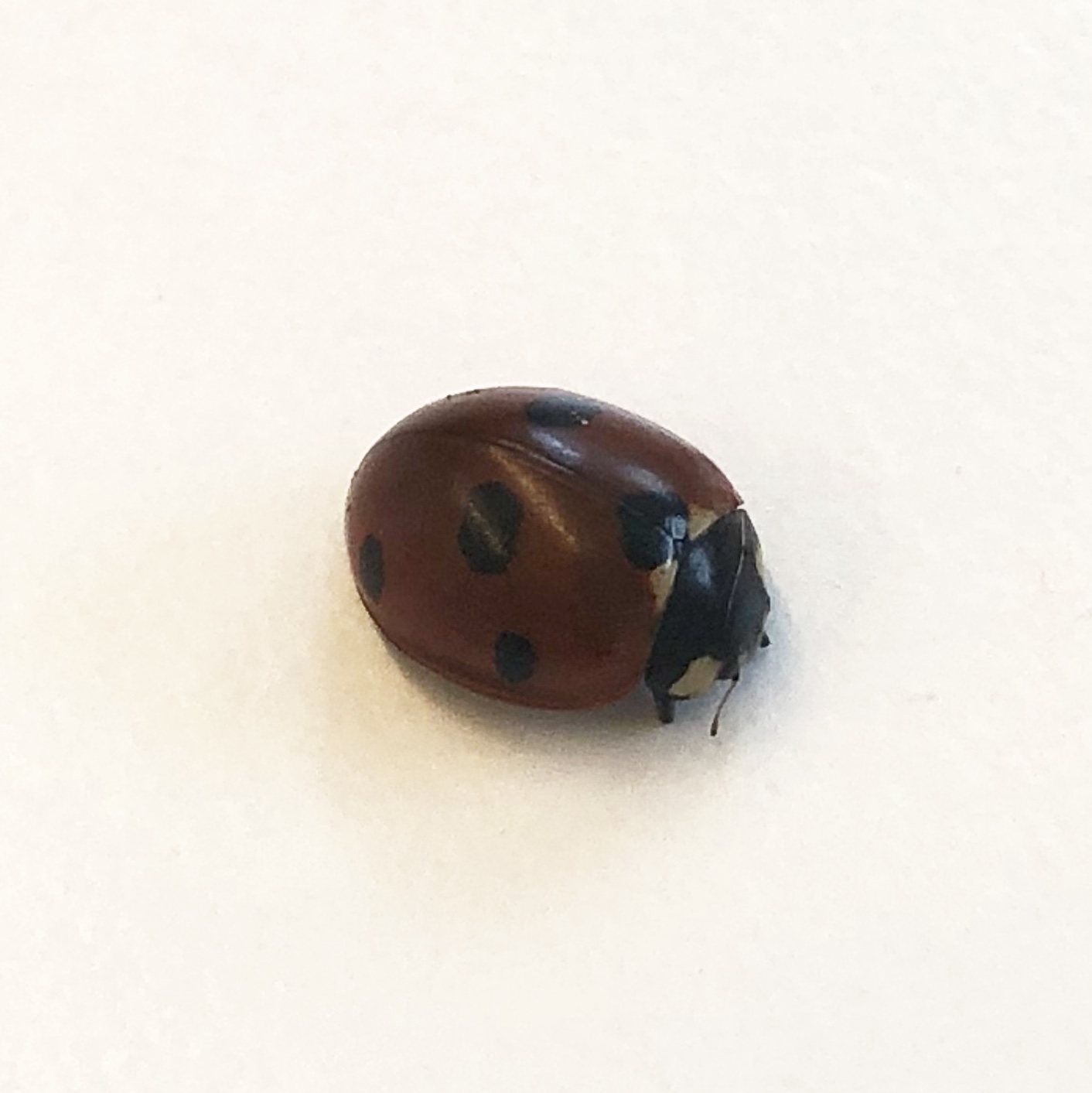}
\centering
\caption{The object, ladybug.}
\label{fig:ladybug}
\end{figure}

\clearpage
\section{Contents of the\\ data set}\label{sec:datasets}

The data set contains the following
CT data files:\\ 
{\tt sinogram1.mat},\\
{\tt sinogram2.mat},\\
{\tt sinogram3.mat},\\
{\tt sinogram4.mat},\\
{\tt sinogram5.mat},\\
{\tt sinogram6.mat},\\
{\tt sinogram7.mat},\\
{\tt sinogram8.mat}\\

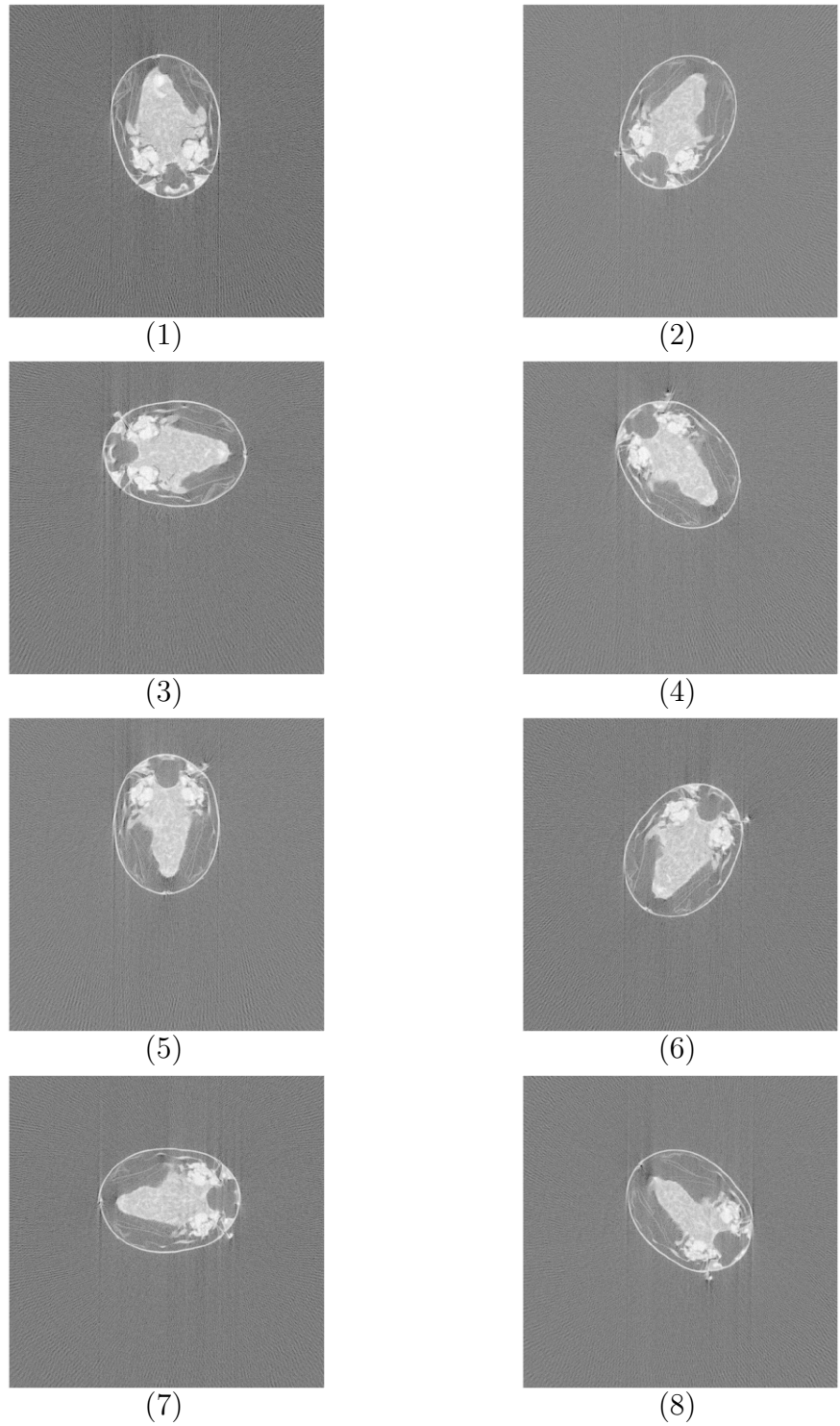
\begin{figure*}
\begin{picture}(390,500)
\put(0,440){\includegraphics[width=200pt]{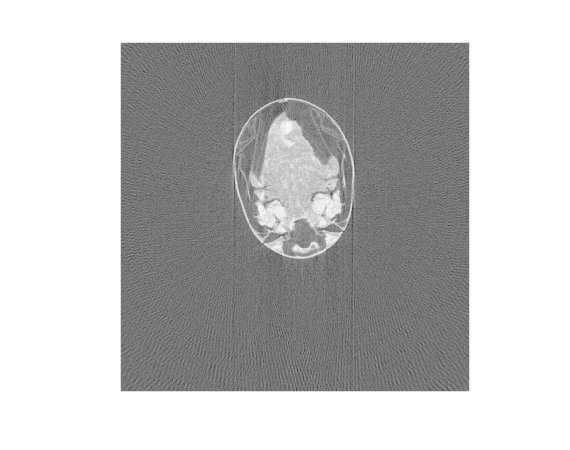}}
\put(200,440){\includegraphics[width=200pt]{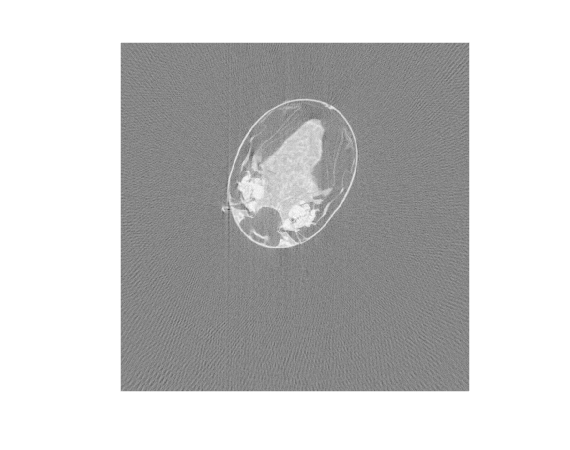}}
\put(0,300){\includegraphics[width=200pt]{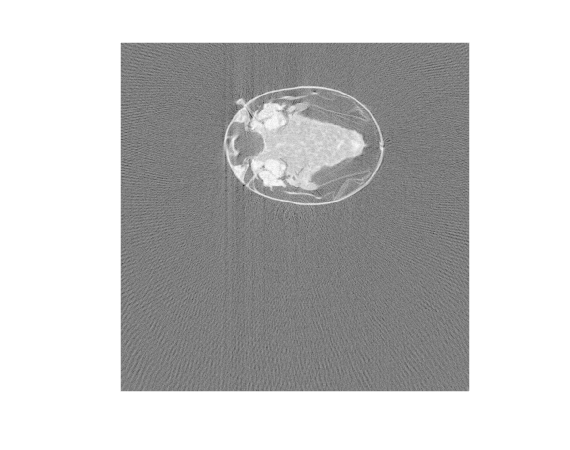}}
\put(200,300){\includegraphics[width=200pt]{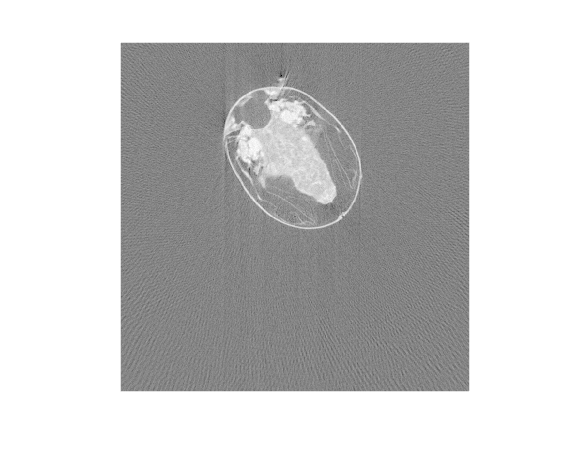}}
\put(0,160){\includegraphics[width=200pt]{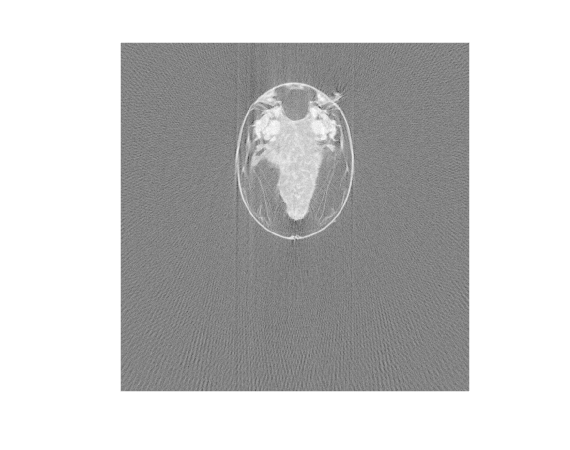}}
\put(200,160){\includegraphics[width=200pt]{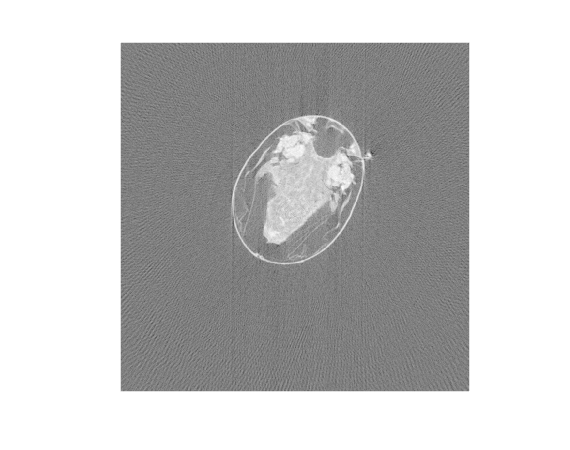}}
\put(0,20){\includegraphics[width=200pt]{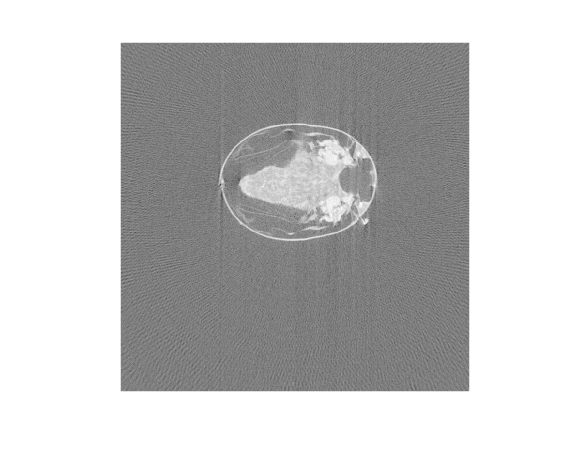}}
\put(200,20){\includegraphics[width=200pt]{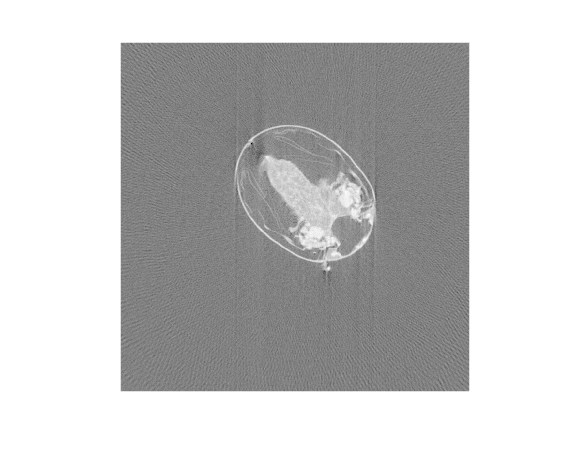}}
\put(95,450){(1)}
\put(295,450){(2)}
\put(95,310){(3)}
\put(295,310){(4)}
\put(95,170){(5)}
\put(295,170){(6)}
\put(95,30){(7)}
\put(295,30){(8)}
\end{picture}
\caption{2D slice (middle slice 483) recontructions from eight different starting positions of the ladybug, with 360 projections using filtered backprojection. Please mention, these pictures are only for visualisation of different positions.}\label{fig:slices}
\end{figure*}

\noindent
The files above include the 3D cone-beam sinograms (1024 x 360 x 186, where 1024 is the width of detector, 360 is the number of projections used in the measurement and 186 denotes the number of rows chosen from the 967 rows of measured data),  
for 8 different measurements starting from different positions (shown in Figure \ref{fig:slices}). Sinograms are taken logarithm and normalized.
From the 2D projection images, different rows corresponding to the central horizontal cross-sections of the ladybug target were taken to form a
cone-beam sinograms.

3D measurement matrices can be freely built for example using the ASTRA \cite{vanAarle2015,vanAarle2016} and opTomo toolbox for MATLAB\footnote{MATLAB is a registered trademark of The MathWorks, Inc.} when more details, for example, the measurement geometry is described in Section \ref{sec:Measurements} below.
The projection angles were the same in every time step.

\bigskip
\bigskip
\bigskip
\bigskip
\bigskip
\bigskip
\noindent
The model for the CT problem is 
\begin{equation}\label{eqn:Axm}
 {\tt Ax=m}, 
\end{equation}
where {\tt A} is the 3D measurement matrix, {\tt m} is the measurement data and {\tt x} is the reconstruction in vector form \cite{kak2002principles,buzug2011computed,mueller2012linear}.
For the dynamic model, a stack of measurement matrices and measurement data sets can be built easily \cite{zang2018space}.

\clearpage

\clearpage
\section{Measurements}\label{sec:Measurements}
\begin{figure}[b]
\includegraphics[width=5cm]{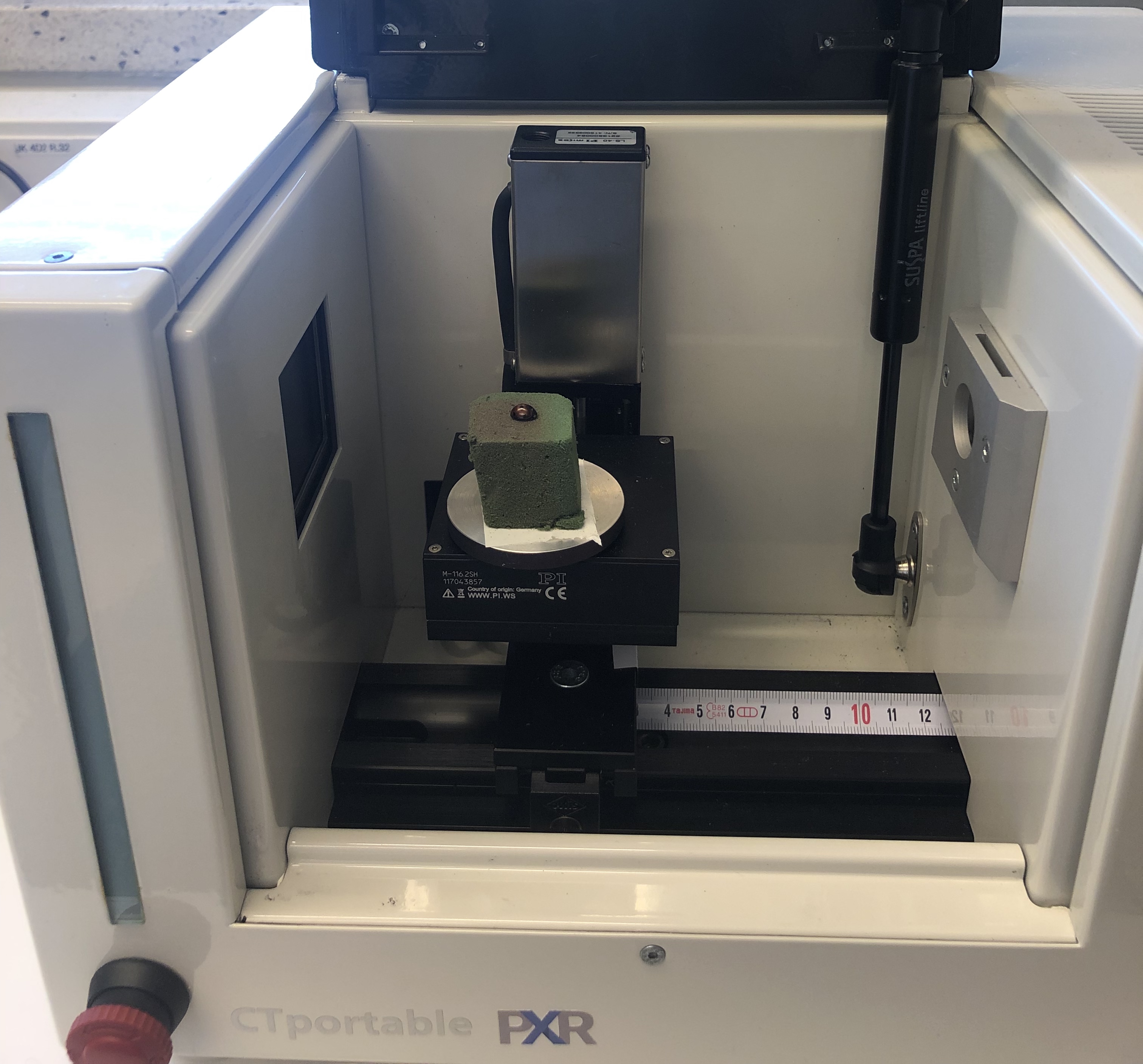}
\centering
\caption{The Procon X-ray CTportable measurement device at Aalto University.}
\label{fig:CTmachine}
\end{figure}

The data in the sinograms are X-ray tomographic (CT) data of a 3D cross-section of the ladybug measured with Procon X-ray CT portable device shown in 
Figure \ref{fig:CTmachine}.

\begin{figure}[h]
\includegraphics[width=7cm]{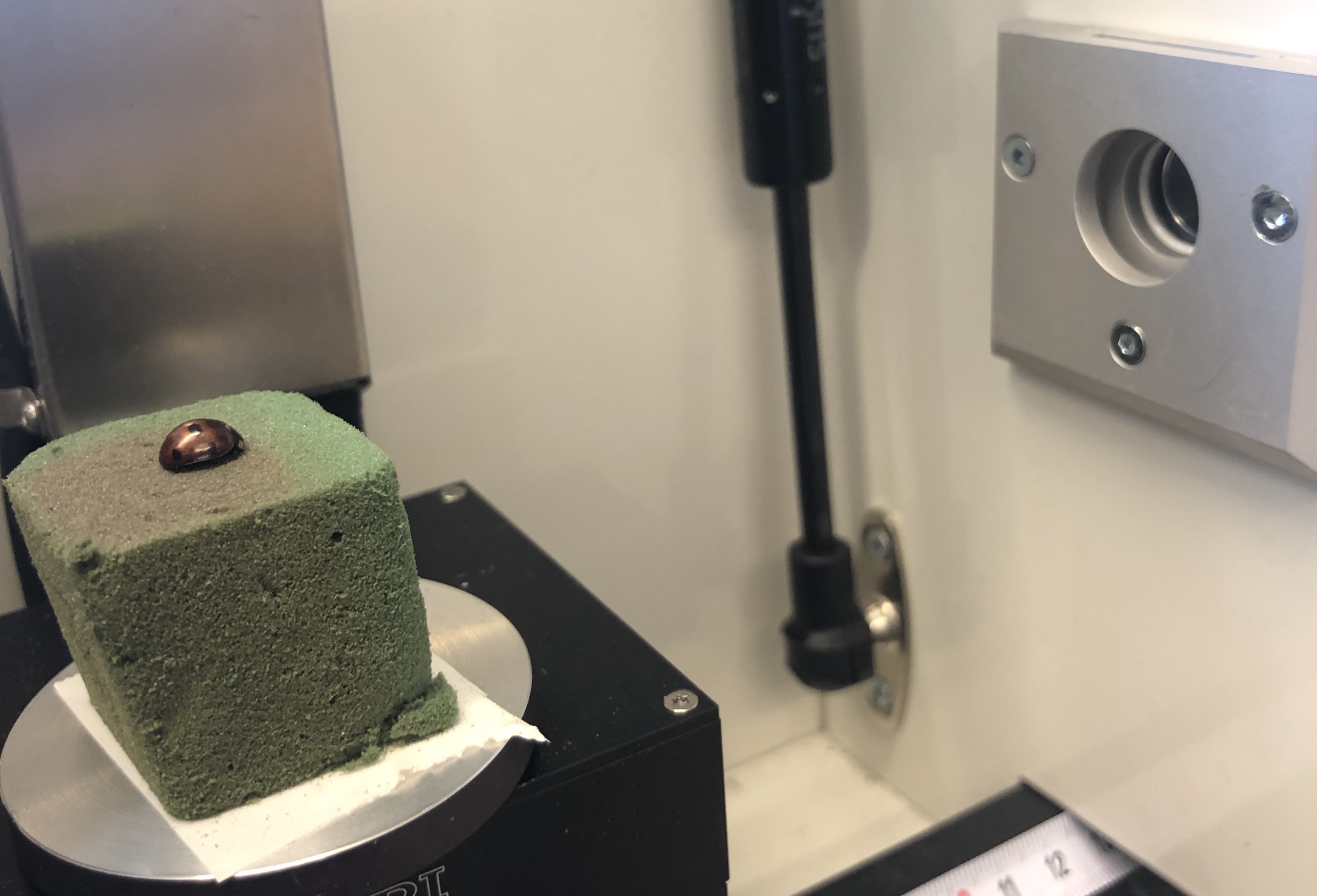}
\centering
\caption{The Procon X-ray CTportable measurement device source.}
\label{fig:CTsource}
\end{figure}

\begin{figure}[h]
\includegraphics[width=5cm]{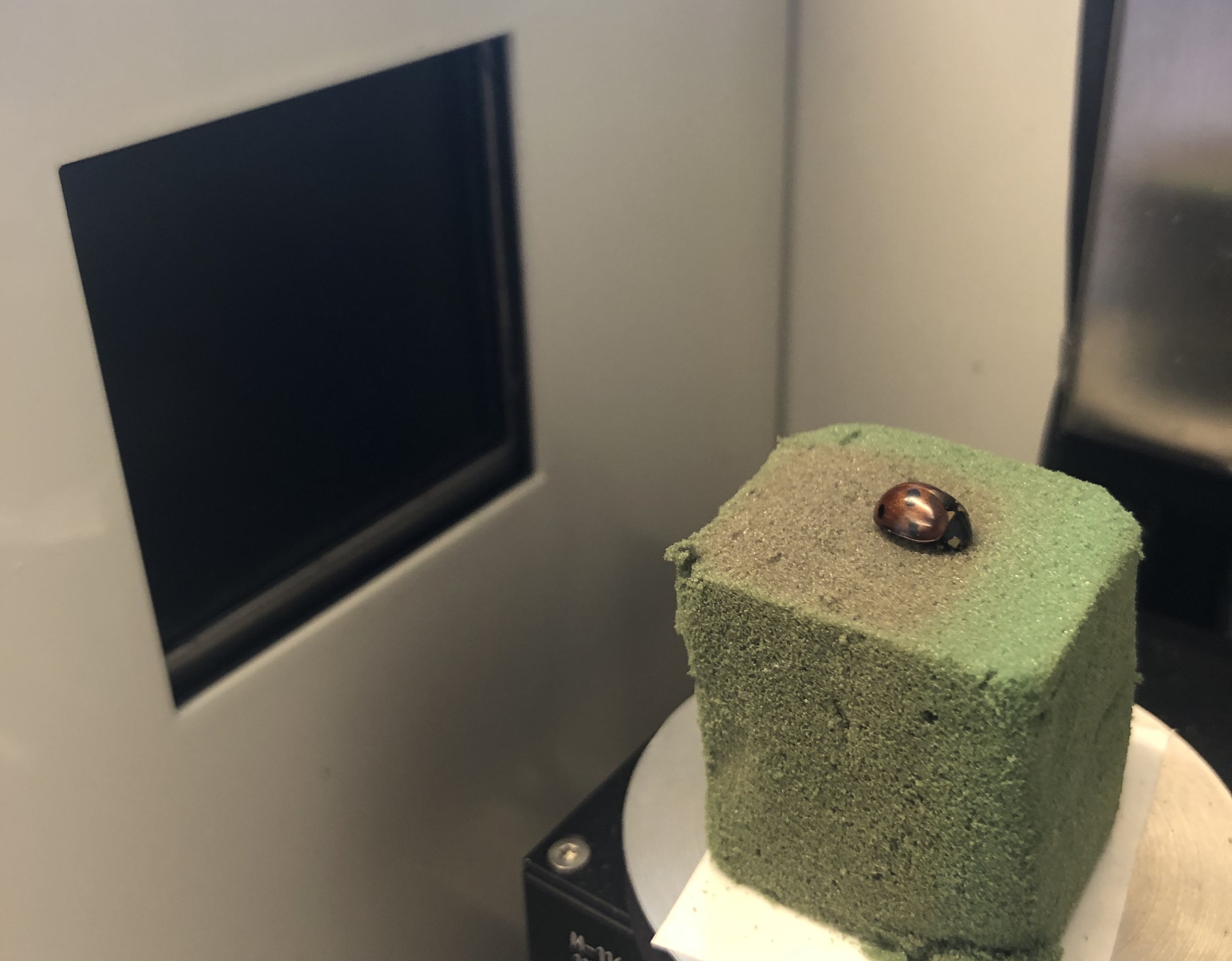}
\centering
\caption{The Procon X-ray CTportable measurement device detector.}
\label{fig:CTdetector}
\end{figure}

\begin{itemize}
\item The X-ray source has 50 kV voltage and 50 W maximum output. The source is shown closely in Figure \ref{fig:CTsource}. 
\item The detector has 1 MP pixels and the pixel size is 48 $\mu$m. The image files generated by the camera were $1024 \times 967$ pixels in size. The detector is shown in Figure \ref{fig:CTdetector}. 
\end{itemize}

\begin{figure}[h]
\includegraphics[width=6cm]{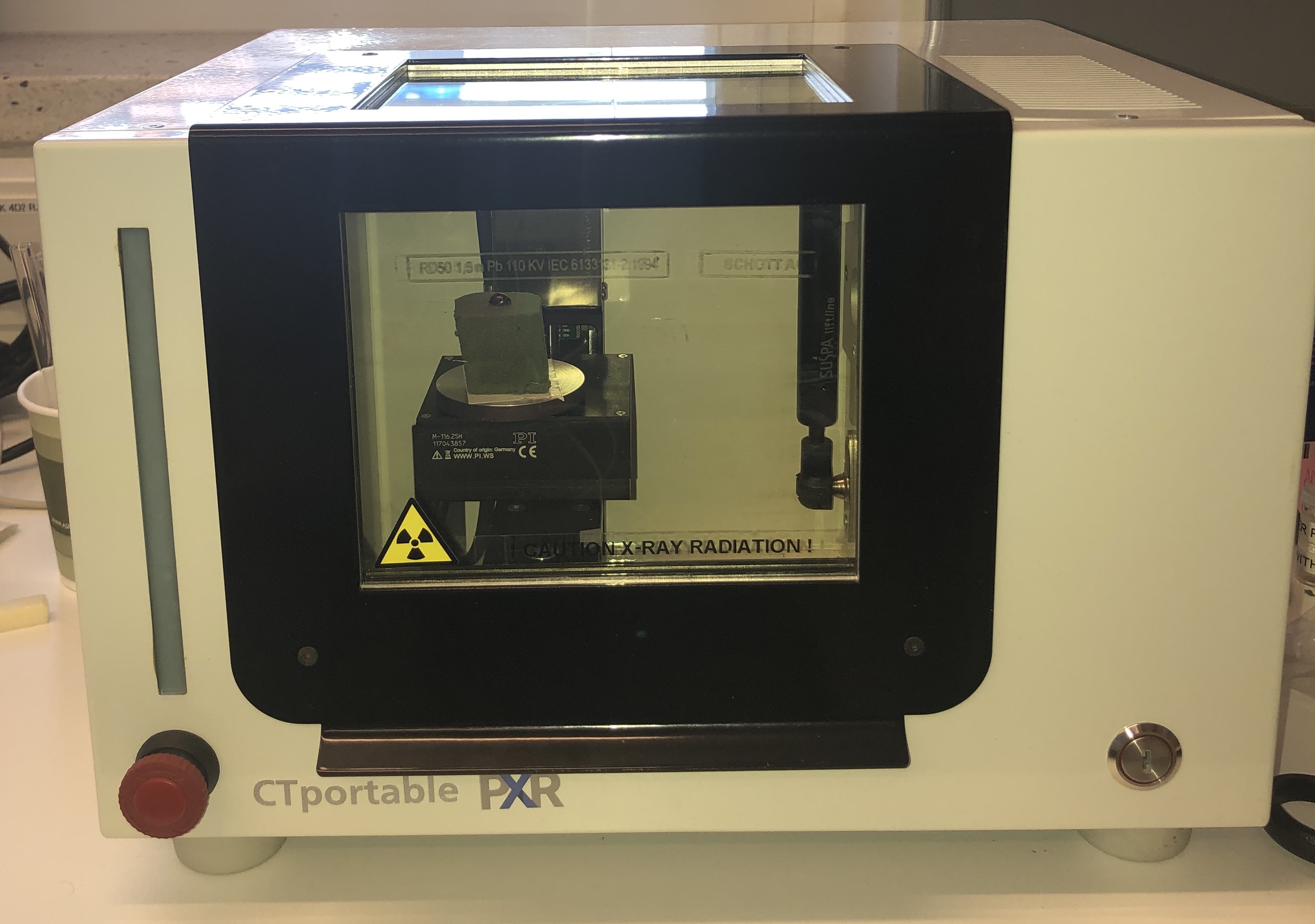}
\centering
\end{figure}

\begin{figure}[h]
\includegraphics[width=6cm]{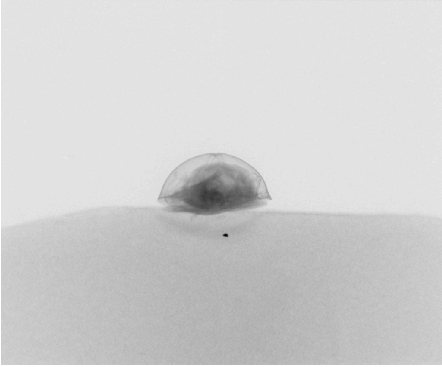}
\centering
\caption{Resulting projection image of the ladybug.}
\label{fig:CTprojection}
\end{figure}

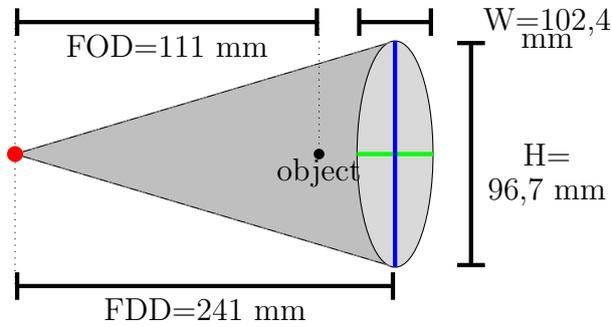
\begin{figure}[h]
\begin{tikzpicture}[rotate = -90, scale=1.0]
\def\rx{1.5}    
\def\ry{0.5}  
\def\z{2.5}     

\pgfmathparse{asin(\ry/\z)}
\let\angle\pgfmathresult

\coordinate (h) at (0, \z);
\coordinate (O) at (0, -2.5);     
\coordinate (A) at ({-\rx*cos(\angle)}, {\z-\ry*sin(\angle)});
\coordinate (B) at ({\rx*cos(\angle)}, {\z-\ry*sin(\angle)});

\draw[fill=gray!50] (A) -- (O) -- (B) -- cycle;
\draw[fill=gray!30] (h) ellipse ({\rx} and {\ry});

\draw [ultra thick,green] (0,2.0) -- (0,3);
\draw [ultra thick,blue] (-1.5,2.5) -- (1.5,2.5);
\draw[|-|, ultra thick] (-1.75,-2.5) -- (-1.75,1.5) node[below,midway]{FOD=111 mm};
\draw [|-|, ultra thick] (1.75,-2.5) -- (1.75,2.5)  node[below,midway]{FDD=241 mm}; 
\draw[|-|, ultra thick] (-1.5,3.5) -- (1.5,3.5); 
\draw[|-|, ultra thick] (-1.75,2) -- (-1.75,3);
\draw[dotted] (-1.75,-2.5) -- (1.75,-2.5); 
\draw[dotted] (-1.75,1.5) -- (0.0,1.5); 
\draw[dashed,gray] (0.0,-2.5) -- (-1.5,2.5);
\draw[dashed,gray] (0.0,-2.5) -- (1.5,2.5); 
\draw (0.0,4.5) node{H=}; 
\draw (0.5,4.5) node{96,7 mm};
\draw (-1.75,4.5) node{W=102,4}; 
\draw (-1.5,4.5) node{mm};
\fill[thick] (0.0,1.5) circle (2pt);
\draw (0.25,1.5) node{object};
\fill[thick, red] (0,-2.5) circle (3pt);
\end{tikzpicture}
\bigskip
\caption{Geometry of the measurement setup.
Here FOD and FDD denote the focus-to-object distance and the focus-to-detector distance, respectively; the black dot object is the center-of-rotation. The height of the detector (the blue line) is denoted by H. The W (and the green vertical line) denotes the width  of the detector. The red dot is the X-ray source. To increase clarity, the $x$-axis and $y$-axis in this image are not in scale.}\label{geometry}
\end{figure}

\clearpage\noindent
The measurement geometry is shown in Figure \ref{geometry}. A set of 360 cone-beam projections with resolution $1024 \times 967$ 
was measured. The exposure time was 400 ms, X-ray tube acceleration voltage 50 kV and tube current 400 mA. See Figure~\ref{fig:CTprojection} for an example of the resulting projection images. 

\begin{figure}[h]
\begin{picture}(100,270)
\begin{tikzpicture}
\put(50,330){\color{gray}\line(1.5,-3){75}}
\put(50,330){\color{gray}\line(-1.5,-3){75}}

\put(50,325){\color{red}\circle*{20}}
\put(0,105){\line(1,0){100}}
\put(0,130){\line(1,0){100}}
\put(0,155){\line(1,0){100}}
\put(0,180){\line(1,0){100}}
\put(0,205){\line(1,0){100}}
\put(0,105){\line(0,1){100}}
\put(25,105){\line(0,1){100}}
\put(50,105){\line(0,1){100}}
\put(75,105){\line(0,1){100}}
\put(100,105){\line(0,1){100}}

\put(125,130){\line(0,1){100}}
\put(106,111){\line(0,1){100}}
\put(113,118){\line(0,1){100}}
\put(119,124){\line(0,1){100}}
\put(25,230){\line(1,0){100}}
\put(6,211){\line(1,0){100}}
\put(13,218){\line(1,0){100}}
\put(19,224){\line(1,0){100}}

\put(4,190){$x_{1,1}$}
\put(4,165){$x_{1,2}$}
\put(10,140){$\vdots$}
\put(2,115){$x_{1,M}$}

\put(28,190){$x_{2,1}$}
\put(28,165){$x_{2,2}$}
\put(35,140){$\vdots$}
\put(26,115){$x_{2,M}$}

\put(55,190){$\cdots$}
\put(55,165){$\cdots$}
\put(55,140){$\ddots$}
\put(55,115){$\cdots$}

\put(78,190){$x_{N,1}$}
\put(78,165){$x_{N,2}$}
\put(85,140){$\vdots$}
\put(76,115){$x_{NM}$}

\put(-25,0){\color{blue}\line(1,0){150}}
\put(-25,25){\color{blue}\line(1,0){150}}
\put(-25,50){\color{blue}\line(1,0){150}}
\put(-25,100){\color{blue}\line(1,0){150}}
\put(-25,75){\color{blue}\line(1,0){150}}

\put(-25,0){\color{blue}\line(0,1){100}}
\put(0,0){\color{blue}\line(0,1){100}}
\put(25,0){\color{blue}\line(0,1){100}}
\put(100,0){\color{blue}\line(0,1){100}}
\put(125,0){\color{blue}\line(0,1){100}}
\put(-23,85){$m_{1,1}$}
\put(2,85){$m_{2,1}$}
\put(30,85){$\cdots$}
\put(83,85){$\cdots$}
\put(101,85){$m_{N,1}$}

\put(-23,60){$m_{1,2}$}
\put(2,60){$m_{2,2}$}
\put(30,60){$\cdots$}
\put(83,60){$\cdots$}
\put(101,60){$m_{N,2}$}

\put(-15,30){$\vdots$}
\put(110,30){$\vdots$}

\put(-24,10){$m_{1,K}$}
\put(1,10){$m_{2,K}$}
\put(30,10){$\cdots$}
\put(83,10){$\cdots$}
\put(100,10){$m_{NK}$}

\end{tikzpicture}
\end{picture}
\caption{The organization of the pixels in the sinograms {\tt m}\,=\,$[m_1,\ldots,m_{N*K}]^T$ and reconstructions {\tt x}\,=\,$[x_1,\ldots,x_{N*M}]^T$ with $N=1024$, $M=360$ and $K=186$. 
The picture shows the organization for the first projection; after that in the full angular view case, the target takes $1$ degree steps 
counter-clockwise (or equivalently the source and detector take steps clockwise) and the following matrices of {\tt m} are determined in an analogous manner.}\label{fig:pixelDemo}
\end{figure}
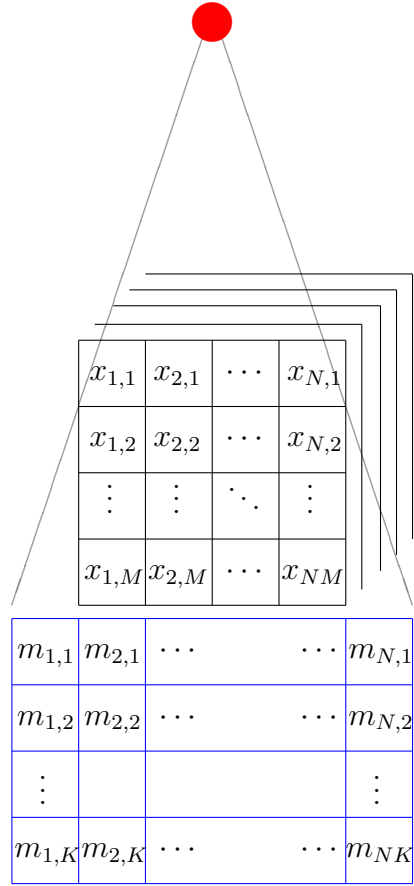

The organization of the pixels in the sinograms and the reconstructions are illustrated in Figure~\ref{fig:pixelDemo}.

\clearpage
\section*{Acknowledgement}
This work was funded by the Academy of Finland and by the School of Electrical Engineering, Aalto University, Finland.

\bibliographystyle{ieeetr}
\bibliography{ref}

\end{document}